\documentclass[a4paper,11pt,aps,superscriptaddress]{revtex4}

\usepackage{amsmath}
\usepackage{graphicx}

\newcommand{\rme}{\ensuremath{\mathrm{e}}}
\newcommand{\D}{\ensuremath{\mathrm{d}}}

\renewcommand{\Re}{\ensuremath{\mathrm{Re}}}
\renewcommand{\Im}{\ensuremath{\mathrm{Im}}}

\renewcommand{\phi}{\ensuremath{\varphi}}
\newcommand{\Cee}{\ensuremath{\mathcal{C}}}
\newcommand{\W}{\ensuremath{\mathcal{W}}}
\newcommand{\Zed}{\ensuremath{\mathcal{Z}}}

\renewcommand{\th}{\mathrm{th}}

\begin{document}

\title{The Lee-Yang theory of equilibrium and nonequilibrium phase transitions}

\date{4th April, 2003}

\author{R.\ A.\ Blythe}
\affiliation{Department of Physics and Astronomy, University of Manchester,
Manchester M13 9PL, UK}

\author{M.\ R.\ Evans}
\affiliation{School of Physics, University of Edinburgh,
Mayfield Road, Edinburgh EH9 3JZ, UK}

\begin{abstract}
We present a pedagogical account of the Lee-Yang theory of equilibrium
phase transitions and review recent advances in applying this theory
to nonequilibrium systems.  Through both general considerations and
explicit studies of specific models, we show that the Lee-Yang
approach can be used to locate and classify phase transitions in
nonequilibrium steady states.
\end{abstract}

\maketitle

\section{Introduction}

In this work we seek a mathematical understanding of phase transitions
in the steady state of stochastic many-body systems.  Systems at
equilibrium with their environment provide examples of such steady
states, and the mechanisms underlying equilibrium phase transitions
are long known and understood.  Experimentally, one can distinguish
between two types of transition: the first-order transition at which
there is phase coexistence, e.g.\ between a high-density solid and a
low-density fluid, and the continuous transition at which fluctuations
and correlations grow to such an extent as to be macroscopically
observable.

From a thermodynamical perspective, one can understand first-order
transitions by associating with each phase a free energy.  For a given
set of external parameters, the phase `chosen' by the system is that
with the lowest free energy, and so a phase transition occurs when the
free energies of two (or more) phases are equal.  The sudden changes
in macroscopically measurable quantities that take place at
first-order transitions are described mathematically as
discontinuities in the first derivative of the free energy.
Discontinuities in higher derivatives relate to continuous
(higher-order) phase transitions.

The tools of equilibrium statistical physics allow one---in principle
at least---to express the free energy solely in terms of microscopic
interactions.  More specifically, the free energy is given by the
logarithm of a \textit{partition function}, a quantity that normalises
the steady-state probability distribution of microscopic
configurations.  Initially it was not universally accepted that this
approach could faithfully describe phase transitions, in particular
the first-order solid-fluid transition \cite{Domb96}.  In order to
show that the statistical mechanical approach \textit{can} reproduce
the correct discontinuities in the free energy at a first-order
transition, Lee and Yang \cite{YL52} introduced a description of phase
transitions concerning zeros of the partition function when
generalised to the complex plane of an intensive thermodynamic
quantity.  Initially, the theory was couched in terms of zeros in the
complex fugacity plane which is appropriate for fluids in the grand
canonical (fluctuating particle number) ensemble.  However, by mapping
a lattice gas onto the Ising model \cite{LY52}, the theory was found
to hold equally well for a magnet in a fixed magnetic field.  Later
\cite{Fisher65,GR69,GL69} it became clear that the distribution of
zeros in the complex temperature plane can reveal information about
phase transitions in the canonical ensemble.

In section~\ref{eqm} we present a brief, self-contained discussion of
the Lee-Yang theory of equilibrium phase transitions which relates
nonanalytic behaviour in the free energy to zeros of the partition
function.  In the remainder of this article, we move our attention to
phase transitions in \textit{nonequilibrium} steady states, that is,
those that carry currents of mass, energy or some other quantity.  We
give therefore a very brief introduction to the established practice
of modelling nonequilibrium physics through stochastic dynamics in
section~\ref{NSS}.  In particular, we will identify a candidate
quantity for use in place of an equilibrium partition function when
performing an Lee-Yang analysis of phase transitions.  Finally, in
section~\ref{neqm} we review recent work in which the Lee-Yang theory
has been successfully applied to specific nonequilibrium phase
transitions.

\section{Overview of Lee-Yang theory of equilibrium phase transitions}
\label{eqm}

In this section we recount the relationship of zeros of a partition
function to phase transitions in the system it describes.  We present
here a brief but self-contained tour of the key points.  For more
detail (and mathematical rigour) the reader is referred to textbooks,
such as \cite{Reichl80} or, for a more advanced presentation,
\cite{Ruelle69}, as well as early treatments of the subject
\cite{YL52,LY52,Fisher65,GR69,GL69}.

For concreteness we shall describe the theory in terms of a simple
model system of $N$ spins in thermal contact with a heat reservoir.
In this model system the energy of the system can take values $E=n
\epsilon$ with $n=0, 1, 2, \ldots , M$ and the number of microstates
corresponding to the $n^\th$ energy level is $g(n)$.  The canonical
partition function is given by
\begin{equation}
\label{Zspinsys}
Z_N(\beta) = \sum_{n=0}^{M} g(n) \exp(-\beta n \epsilon)
\end{equation}
where $\beta$ is the inverse temperature.  To
investigate the zeros of this partition function, it is useful to make
the change of variable $z = \rme^{-\beta \epsilon}$. Then, the
partition function is explicitly a polynomial in $z$ of degree $M$ and
can be factorised as
\begin{equation}
\label{Zfactor}
Z_N(z) = \sum_{n=0}^{M} g(n) z^n = \kappa \prod_{n=1}^{M} \left( 1 -
\frac{z}{z_n} \right) \;.
\end{equation}
Clearly the quantities $z_n$ are the $M$ zeros of the partition
function $Z_N(z)$; meanwhile, $\kappa$ is a constant which we can
safely ignore in the following.

Since all the $g(n)$ are positive, no zero of $Z_N(z)$ can be real and
positive: that is, the zeros $z_n$ will generally lie in the complex
plane away from the physical values of $z$ which lie on the positive
real axis.  We now define for all complex $z$ \textit{except} the
points $z=z_n$ the (complex) free energy
\begin{equation}
\label{hdef}
h_N(z) \equiv \frac{\ln Z_N(z)}{N} \;.
\end{equation}
Using (\ref{Zfactor}) we rewrite this as
\begin{equation}
\label{ffactor}
h_N(z) = \frac{1}{N} \sum_{n=1}^{M} \ln \left( 1 - \frac{z}{z_n} \right)
\end{equation}
and note that a Taylor series expansion of $h_N(z)$ around a point
$z\ne z_n$ has a finite radius of convergence given by the distance to
the nearest zero from $z$.  This then implies that that $h_N(z)$ can
be differentiated infinitely many times over any region of the complex
plane that is devoid of partition function zeros.  Since we identify a
phase transition through a discontinuity in a derivative of the free
energy, we see that such a transition can only occur at a point $z_0$
in the complex plane if there is at least one zero of the partition
function $Z_N(z)$ within any arbitrarily small region around the point
$z_0$.

Clearly this scenario is impossible if the number of zeros $M$ is
finite, except at the isolated points $z_n$ where the free energy
exhibits a logarithmic singularity.  Since such a point cannot lie on
the positive real $z$ axis, there is no scope for a phase transition
in a finite spin system, such as the simple example (\ref{Zspinsys}).
On the other hand, if the partition function zeros accumulate towards
a point $z_0$ on the real axis as we increase the number of spins $N$
to infinity there is the possibility of a phase transition.

In order to deal with the thermodynamic limit (see \cite{Ruelle69} for
rigorous considerations) we shall assume that the limit
\begin{equation}
h(z) = \lim_{N\to\infty} \frac{\ln Z_N(z)}{N}
\end{equation}
exists and we may write
\begin{equation}
h(z) = \int \D z^\prime \rho(z^\prime) \ln \left( 1 - \frac{z}{z'}
\right)
\end{equation}
where $\rho(z)$ is the local density of zeros in the complex-$z$ plane
in the thermodynamic limit.

Since the imaginary part of this complex free energy $h(z)$ is
multi-valued it will at times be more convenient to work with the
\textit{potential} $\phi(z)$ defined as
\begin{equation}
\label{potential}
\phi(z) \equiv \Re\, h(z) = \int \D z^\prime \rho(z^\prime) \ln \left| 1 -
\frac{z}{z'} \right| \;.
\end{equation}

An expression for the density $\rho(z)$ in terms of the potential
$\phi(z)$ is easily obtained once one recognises that $\ln|z|$ is the
Green function for the two-dimensional Laplacian.  Specifically, if
$z=x+iy$, we have
\begin{equation}
\nabla^2 \ln |z| \equiv \left( \frac{\partial^2}{\partial x^2} +
\frac{\partial^2}{\partial y^2} \right) \ln | x+iy | = 2\pi \delta(x)
\delta(y) \;.
\end{equation}
Using this expression, we can take the Laplacian of both sides of
(\ref{potential}) to find
\begin{equation}
\label{subharmonic}
\rho(z) = \frac{1}{2\pi} \nabla^2 \phi(z) \;.
\end{equation}
Such an equation is familiar from electrostatics and so $\phi(z)$ is
analogous to an electrostatic potential.

In analogy to electrostatics, as long as we can integrate $\rho(x)$
over  bounded regions containing any singularities of $\rho$, the
potential will be a continuous function \cite{kellogg}.  The
significance of this statement is that we can derive a rule for
locating phase boundaries given a partition function.  Let us suppose
that around points $z_1$ and $z_2$ in the complex plane, one has
(analytic) expressions for the potential $\phi_1(z)$ and $\phi_2(z)$
such that $\phi_1(z) \not\equiv \phi_2(z)$ (i.e.\ not the same
function over the entirety of the complex plane).  In order for the
potential to be continuous at all points on the complex plane, we must
have a phase boundary at those values of $z$ for which the condition
\begin{equation}
\label{boundary-rule}
\phi_1(z) = \phi_2(z)
\end{equation}
holds.  This is basically the definition of a transition mentioned
above in the introduction.  Since $\phi_1$ and $\phi_2$ are different
functions, some derivatives of $\phi(z)$ will not exist at these
values of $z$ and we expect the density of zeros at these points to be
non-zero.  (It is also possible for zeros to be present at other
points in the complex plane, but we do not need to consider this
possibility here.)

Typically, a solution of (\ref{boundary-rule}) describes a curve $C$
that intersects the positive real $z$ axis at a point $z_0$.  Having
already established that we require zeros to accumulate at the point
$z_0$ in the thermodynamic limit for a physical phase transition to
take place, we are interested in the line density of zeros per unit
length of this curve.  We introduce the arclength $s$ which measures
the distance along $C$ from the transition point $z_0$.

\begin{figure}
\begin{center}
\includegraphics[scale=0.6]{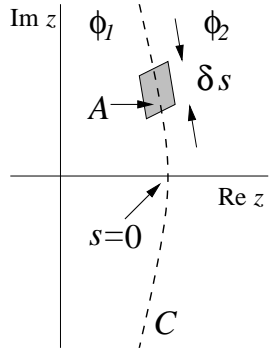}
\end{center}
\caption{\label{line-density} Small area $A$ covering a length $\delta
s$ of the curve of zeros $C$ required to derive a relationship for the
line density $\mu(s)$ of zeros along the curve.}
\end{figure}

To obtain an expression for the line density of zeros $\mu(s)$ along
the curve $C$, consider a short section of $C$ of length $\delta s$
enclosed by an area $A$ that has two sides parallel to $C$ and the
other two sides perpendicular to $C$---see figure~\ref{line-density}.
Integrating the density $\rho(z)$ over this area we have
\begin{equation}
\label{density-integral1}
\int_A \D z \rho(z) = \mu(s) \delta s \;.
\end{equation}
Meanwhile, we have from (\ref{subharmonic}) and an application of the
divergence theorem that
\begin{equation}
\label{density-integral2}
\int_A \D z \rho(z) = \frac{1}{2\pi} \int_A \D z \nabla \cdot [\nabla
\phi (z)] = \frac{\delta s}{2\pi} [\nabla \phi_2(z) - \nabla
\phi_1(z)] \cdot \hat{n}
\end{equation}
in which the functions $\phi_1(z)$ and $\phi_2(z)$ relate to the
limiting values of $\phi(z)$ as the point $z$ on the curve $C$ is
approached from either side, and the vector $\hat{n}$ is the unit
vector normal to $C$ at that point.

Recall that $\phi(z)$ is the real part of the complex free energy
$h(z)$ and therefore, away from any zeros of the partition function,
satisfies the Cauchy-Riemann equations
\begin{equation}
\partial_x \phi(z) = \partial_y \psi(z) \quad \mbox{and} \quad
\partial_y \phi(z) = - \partial_x \psi(z)
\end{equation}
in which $\psi(z)$ is the imaginary part of $h(z)$ and $z=x+iy$.  Then
we have that
\begin{equation}
\nabla \phi(z) \cdot \hat{n} = \nabla \psi(z) \cdot \hat{t}
\end{equation}
where $\hat{t}$ is the unit vector tangent to the curve $C$ at the
point $z$.  Thus we recognise the scalar product in
(\ref{density-integral2}) as the directional derivative of the
imaginary part of the free energy $h(z)$ along the curve $C$.  Putting
this together with (\ref{density-integral1}) we find that
\begin{equation}
\label{density-rule}
\mu(s) = \frac{1}{2\pi} \frac{\D}{\D s} [ \psi_2(z) - \psi_1(z) ]
\end{equation}
in which $\psi(z) = \Im [h(z)]$ and the subscripts indicate opposite
sides of a phase boundary.

Let us now assume that in the thermodynamic limit, there is a phase
transition at the point $z_0$.  Then, either side of the transition
the free energy can be written as
\begin{equation}
f_{1,2}(\tilde{z}) = h(z_0) + b_{1,2} \tilde{z} + c_{1,2} \tilde{z}^2
+ \ldots
\end{equation}
where $\tilde{z}=z-z_0$ and $f_1(\tilde{z})$ is valid for $\Re\,
\tilde{z} < 0$ and $f_2(\tilde{z})$ valid for $\Re\, \tilde{z} >
0$. Note that for the free energy to be real along the real $z$ axis,
the coefficients $b$ and $c$ must also be real.  From the criterion
that the real part of $h(z)$ is continuous across a phase boundary, we
find that the boundary lies along the curve
\begin{equation}
\tilde{y}^2 = \tilde{x}^2 + \frac{b_2-b_1}{c_2-c_1} \tilde{x}
\end{equation}
where $\tilde{x}$ and $\tilde{y}$ are the real and imaginary parts of
$\tilde{z}$ respectively. We consider the conditions under which a
discontinuous (first order) or continuous (second or higher order)
appear.
\vspace*{1ex}

\noindent{\bf First-order transition}\\
For the case where $b_1 \ne b_2$, and the free energy has a
discontinuity in its first derivative, the curve of zeros is a
hyperbola that passes smoothly through the transition point $z_0$.
Hence the tangent to the curve of zeros is parallel to the imaginary
axis at $z_0$ and using the rule (\ref{density-rule}) we find that
\begin{equation}
\mu(0) = \frac{1}{2\pi} \left. \frac{\D}{\D \tilde{y}}
[ (b_2-b_1) \tilde{y} + 2 (c_2-c_1) \tilde{x} \tilde{y} ]
\right|_{\tilde{x}=\tilde{y}=0} = \frac{b_2-b_1}{2\pi} \;.
\end{equation}
Hence we see that the density of zeros at the transition point $z_0$
is nonzero at a first-order phase transition (i.e.\ one at which the
first derivative of the free energy is discontinuous).
\vspace*{1ex}

\noindent{\bf Second-order transition}\\
If $b_1=b_2$ but $c_2 \ne c_1$ we have that the curve $C$ obeys the
equation $\tilde{y} = \pm \tilde{x}$.  Since the zeros of the
partition function come in complex-conjugate pairs, we find that the
zeros approach the point $z_0$ along straight lines that meet at a
right-angle.  If we consider the line
$\tilde{x}=\tilde{y}=s/\sqrt{2}$, we find from (\ref{density-rule})
that
\begin{equation}
\mu(s) = \frac{1}{2\pi} \frac{\D}{\D s} (c_2-c_1) s^2 =
\frac{c_2-c_1}{\pi} s \;.
\end{equation}
This reveals that at a second-order phase transition, the density of
partition function zeros decreases linearly to zero at the phase
transition point.
\vspace*{1ex}

\noindent{\bf Higher-order transition}\\
More generally, one can consider a difference in the free energies
either side of the transition point to have the leading behaviour
$f_2(u) - f_1(u) \sim u^\alpha$.  Then the condition that $\Re[f_2(u)
- f_1(u)]=0$ suggests that the curve of zeros approaches the real axis
at an angle $\frac{\pi}{2\alpha}$.  Unless $\alpha=1$, the curve does
not pass smoothly through the real axis.  In any case, the imaginary
part of the free energy difference grows as $|u|^\alpha$ giving rise
to a density of zeros that behaves as $s^{\alpha-1}$ for small
arclength $s$.  This means that for the density of zeros to be finite
at the transition point we must have $\alpha \ge 1$.
\vspace*{1ex}

In this section we have summarised what we refer to as the Lee-Yang
theory of phase transitions.  This describes how partition function
zeros are related to phase transitions: the accumulation of zeros at a
point along the physical (real, positive) axis of the control
parameter gives the critical value and the density of zeros near to
the accumulation point determines the order of the phase transition.

We derived a rule (\ref{boundary-rule}) for locating phase boundaries
and a further rule (\ref{density-rule}) for finding the density of
zeros along such boundaries.  At a first-order transition there is a
nonzero density of zeros at the transition point whereas the density
decays as a power-law to zero at the transition point when the
associated phase transition is continuous.

Although we have discussed the theory of partition function zeros with
reference to the specific system described by the partition function
(\ref{Zspinsys}) we should note that the ideas hold much more
generally.  Firstly, one is not restricted to the canonical ensemble:
indeed in the original exposition of the theory \cite{YL52}, Lee and
Yang worked in the grand canonical ensemble in which the quantity
generalised to the complex plane was not a function of the temperature
but rather the chemical potential.  Of course, in the equilibrium
theory, these intensive field-like quantities play similar
mathematical roles and so there is no reason why the Lee-Yang theory
shouldn't apply to all of them.  For historical reasons, zeros in the
complex fugacity (or chemical potential) plane are often referred to
as Lee-Yang zeros \cite{YL52,LY52} and those in the complex
temperature plane Fisher zeros \cite{Fisher65}.  Also it appears that
one can just as easily generalise physical `field-like' variables
(such as temperature) or `fugacity-like' variables (such as the
quantity $z$ considered above) to the complex plane without altering
the properties of the partition-function zero densities at first-order
and continuous phase transitions described above.

Despite the apparently wide generality of the Lee-Yang theory of
equilibrium phase transitions, proving its validity in the general
case is a difficult task (although we note recent work in this
direction \cite{BBCKK00}).  Therefore, most rigorous results, such as
those discussed in \cite{Ruelle69}, tend to rely on specific
properties of a particular partition function.  Perhaps the most
spectacular of these is that obtained by Lee and Yang in their
original work.  Specifically, they found that the zeros of the
partition function for an Ising ferromagnet in an external field $h$
all lie on the circle $|\exp(h)|=1$.  The significance of this result,
which does not depend on the number of spatial dimensions, lattice
structure and details of the spin-spin interactions, is that if there
is a phase transition induced by varying the magnetic field $h$, it can
only occur at the point $h=0$, and then only if the partition function
zeros accumulate there in the thermodynamic limit.

\section{Nonequilibrium steady states}
\label{NSS}

A nonequilibrium steady state differs from its equilibrium counterpart
in that it may admit a flow of, say, energy or mass.  More generally,
these states have a circulation of probability in the space of
microscopic configurations.  In the past few years, it has become
customary to model nonequilibrium systems as stochastic processes,
i.e.\ simple models defined by local dynamical rules.  Extensive study
of these has revealed a range of phenomena including nonequilibrium
phase transitions
\cite{MD99,Evans00,Hinrichsen00,Mukamel00,Schutz01,Stinchcombe01,EB02}.
We present below the key elements of this approach to nonequilibrium
physics with a view to understanding nonequilibrium phase transitions
in the framework of the Lee-Yang theory described above.  In
particular, we will need to propose a quantity to use in place of the
equilibrium partition function (\ref{Zspinsys}).

Let us begin by discussing how one might realise the dynamics of the
\textit{equilibrium} spin system of the previous section, for example
in a computer simulation.  The aim is to generate a sequence of spin
configurations such that the frequency with which a particular
configuration $\Cee$ is generated is proportional to the Boltzmann
weight $f(\Cee) = \exp[-\beta E(\Cee)]$ where $E(\Cee)$ is the energy
of configuration $\Cee$.  Usually this is achieved by choosing the
next configuration $\Cee^\prime$ from the current configuration $\Cee$
with a probability proportional to the transition rate $W(\Cee \to
\Cee^\prime)$ that satisfies the detailed balance condition
\begin{equation}
\label{db}
f(\Cee) W(\Cee \to \Cee^\prime) =
f(\Cee^\prime) W(\Cee^\prime \to \Cee) \quad \forall\, \Cee, \Cee^\prime \;.
\end{equation}
Apart from the fact that this choice of transition rates guarantees
convergence to the desired equilibrium distribution of microstates
\cite{NY01}, it also has the feature that the mean current of any
quantity in the steady state is zero (as it should be at equilibrium).

Since we are interested in nonequilibrium systems that support
steady-state currents we must work with transition rates that do not
satisfy the condition (\ref{db}) or equivalent criteria (given in
e.g.\ \cite{vanKampen92,Mukamel00,EB02}).  Clearly one has a lot of
freedom in the choice of transition rates, and so in practice one
often devises dynamics that seem physically reasonable given the
phenomena one is trying to describe.  Later, in section~\ref{DDS} we
will give concrete examples in the form of driven diffusive systems,
in which particle moves are biased in a particular direction to model
the effect of an external field.

In order to find the set of steady-state weights associated with a
particular set of prescribed transition rates, we impose the condition
that the total flow of probability into a configuration $\Cee$ is
balanced by the corresponding outflow.  That is, we must have
\begin{equation}
\label{steady-state}
\sum_{\Cee^\prime \ne \Cee} \left[ f(\Cee^\prime) W(\Cee^\prime \to
\Cee) - f(\Cee) W(\Cee \to \Cee^\prime) \right] = 0
\end{equation}
for every configuration $\Cee$.  In general there might be more than
one solution to this set of equations for the rates $f(\Cee)$, each
corresponding to a steady state that is reached with some probability
that depends on the starting configuration.  We shall assume for
simplicity that the steady state is unique and therefore reached with
certainty from every initial state.  A sufficient condition for a
unique steady state is that it is possible to reach each configuration
from every other via a sequence of microscopic transitions
\cite{EB02}.

Once the steady state weights $f(\Cee)$ are known from
(\ref{steady-state}) one can obtain the corresponding probability
distribution of microscopic configurations $P(\Cee)$ through a
normalisation
\begin{equation}
\label{Zneqm}
Z = \sum_{\Cee} f(\Cee)
\end{equation}
such that $P(\Cee) = f(\Cee)/Z$.  Then, one can compute averages of
physical quantities and look for nonanalyticities to locate
nonequilibrium phase transitions as one varies the transition rates.
At equilibrium we saw that the zeros of the function (\ref{Zspinsys})
that normalises the Boltzmann weights encodes the phase behaviour of
the system.  Thus we might hope that more generally, the zeros of the
normalisation (\ref{Zneqm}) will provide information about
nonequilibrium phase behaviour.

In section~\ref{neqm} we review recent work that suggests that the
steady-state phase behaviour of certain nonequilibrium driven diffusive
and reaction-diffusion systems \textit{is} correctly described by the
zeros of the normalisation $Z$.  Although we are unaware of any
rigorous argument for this to be the case, some suggestive evidence is
provided by an observation made in \cite{Blythe01,EB02} which we now
outline.

First note that the equation (\ref{steady-state}) is linear in the
weights $f(\Cee)$.  This implies that these weights, and hence the
normalisation, can always be written as sums of products of the
transition rates $W(\Cee \to \Cee^\prime)$.  Now, by numbering the
microscopic configurations, one can construct the transition rate
matrix $\W$ whose off-diagonal elements $\W_{nm}$ are equal to the
transition rates from configuration $m$ to $n$ and the diagonal
elements $\W_{nn}$ are negative and express the total rate of
departure from configuration $n$.  Using elementary results from
matrix theory (see \cite{Blythe01,EB02} for the details), one can show
that an expression for the normalisation $Z$ that is polynomial in the
transition rates $\W(\Cee\to\Cee^\prime)$ is
\begin{equation}
\label{Zevprod}
Z = \prod_{\lambda_i \ne 0} ( - \lambda_i )
\end{equation}
in which $\lambda_i$ are the eigenvalues of the matrix $\W$.

With each of the eigenvalues $\lambda_i$ is associated an eigenvector
of $\W$ describing a `mode' of the stochastic process that decays
exponentially with a timescale $\tau_i = 1/|\lambda_i|$ \cite{EB02}.  We
have assumed that the process described by $\W$ has only one steady
state, and so only one of the eigenvalues is equal to zero (since
clearly the relaxation time of a steady state is infinite).  Equation
(\ref{Zevprod}) states that the normalisation $Z$ can be written as a
product of the remaining eigenvalues.  Now, at a phase transition we
expect diverging timescales: physically one encounters metastable
(long-lived) states near first-order transition points and long
correlation lengths and times at continuous transitions.  The presence
of long timescales $\tau_i$ implies small eigenvalues $\lambda_i$ of
$\W$ and, from (\ref{Zevprod}), that the normalisation $Z$ approaches
zero.  Hence it appears that it is appropriate to consider the zeros
of $Z$ as given by (\ref{Zneqm}) to locate nonequilibrium phase
transitions.

\section{Application of Lee-Yang theory to nonequilibrium phase transitions}
\label{neqm}

We shall now review recent progress in applying the Lee-Yang theory to
nonequilibrium phase transitions.  We consider first in section
\ref{DDS} driven diffusive systems, where most work has so far been
focussed \cite{Arndt00,BE02, Jafarpour02}.  An appealing feature,
discussed in more detail below, is that some one-dimensional cases
have been solved exactly, the normalisation (\ref{Zevprod})
calculated, and nonequilibrium phase transitions analysed.  (The
existence of one-dimensional phase transitions contrasts with the case
of one-dimensional equilibrium models which do not admit phase
transitions if the interactions are short-ranged.)  Then in Section
\ref{RDandDP} we move onto reaction-diffusion systems and directed
percolation.

\subsection{Driven diffusive systems}
\label{DDS}

In their original papers on partition function zeros, Lee and Yang
\cite{YL52,LY52} made use of the mapping between the Ising model of a
magnet and the lattice gas.  Essentially one associates up-spins with
particles and down-spins with vacancies so that a positive interaction
strength $J$ in the Ising model results in a particle-particle
attraction in the lattice gas, and a negative interaction strength
gives rise to repulsive gas particles.  Note that there is implicitly
at most one particle per lattice site, so there is a hard-core
exclusion in the lattice gas.

As discussed in section~\ref{NSS}, one can realise the dynamics of the
lattice gas through a set of transition rates that satisfy the
detailed balance condition (\ref{db}) with respect to the Boltzmann
distribution.  Thirty years after Lee and Yang's work, Katz, Lebowitz
and Spohn (KLS) \cite{KLS83,KLS84} introduced a \textit{driven}
lattice gas model in which the rate at which particles hop in the
direction of an external field is enhanced and the hop-rate against
the field is suppressed.  This model is well-studied and many results
are discussed in \cite{SZ95}.  The principal effect of the driving
field is to introduce anisotropy in the phase-separation that occurs
below the critical temperature associated with the spontaneous
magnetisation of the Ising ferromagnet ($J>0$) in two or more spatial
dimensions and in the critical exponents that characterise this
transition.

As yet, the KLS model remains unsolved for general interaction
strength, although in one dimension the steady state is known for some
parameters \cite{KS92}.  The particular case of one dimension and zero
interaction strength $J=0$, known as the asymmetric simple exclusion
process (ASEP), had already been studied---at least at a mean-field
level---by biophysicists interested in the kinetics of
biopolymerisation \cite{MGP68}.  The mean-field approach predicts
phase transitions in the steady state as parameters controlling the
rate of insertion and extraction of particles at the boundaries are
varied \cite{DDM92}.  The existence of these phase transitions is
confirmed through an exact solution of the ASEP
\cite{DDM92,DEHP93,SD93}, achieved using a powerful matrix product
approach \cite{DEHP93,Derrida98} which has subsequently been used to
solve many generalisations of the ASEP.  The details of the matrix
product method are not necessary for the following---suffice to say
that one ends up calculating a normalisation proportional to
(\ref{Zevprod}) through a product of matrices, often of infinite
dimension. In this way one obtains some explicit formulas for the
normalisation (\ref{Zneqm}) which we shall use below \footnote{As yet,
there is no comprehensive review of applications of the matrix-product
method, although one should consult
\cite{Derrida98,Hinrichsen00,Schutz01,Stinchcombe01} and references
therein}.

The asymmetric exclusion process with open boundaries is perhaps the
simplest exactly solved nonequilibrium model that exhibits both a
first-order and continuous phase transition in its steady state.
Therefore it is an ideal candidate for testing the hypothesis outlined
in section~\ref{NSS} that zeros of the normalisation should accumulate
towards the positive real axis in the complex plane of transition
rates.  Before outlining the results of this analysis (the details of
which are presented in \cite{BE02}) we recall the definition of the
ASEP with open boundaries.

\begin{figure}
\begin{center}
\includegraphics[scale=0.75]{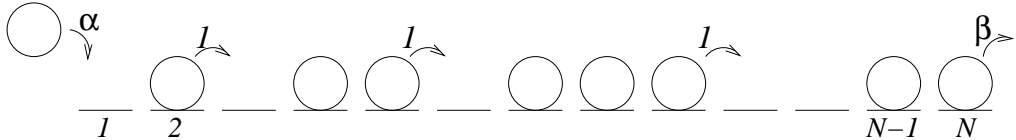}
\end{center}
\caption{\label{ASEP} The dynamics defining the ASEP.  
The lattice is open and particles may be inserted or
extracted at the boundaries as shown.}
\end{figure}

\begin{figure}
\begin{center}
\includegraphics[scale=0.75]{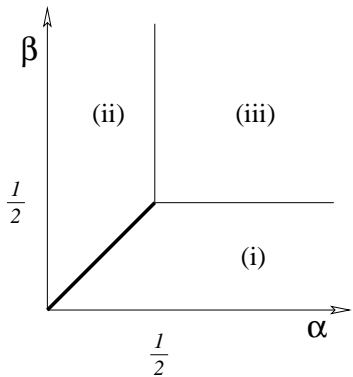}
\end{center}
\caption{\label{ASEP-phase} The phase diagram of the ASEP in the space
spanned by the boundary rates $\alpha$ and $\beta$.  The thick line
represents a first-order phase boundary, the thin line a continuous
phase boundary.}
\end{figure}

In this system, a particle on an $N$-site lattice can hop one site to
the right at unit rate, as long as the receiving site is empty.
Meanwhile, particles are inserted onto the leftmost lattice site (if
empty) at a rate $\alpha$ and removed from the rightmost site (if
occupied) at a rate $\beta$---see figure~\ref{ASEP}.  Along the line
$\alpha = \beta < \frac{1}{2}$ there is a coexistence between a high-
and a low-density phase, with a shock separating the two.  This is
indicative of a first-order transtion.  Meanwhile, along the lines
$\alpha> \frac{1}{2}, \beta=\frac{1}{2}$ and $\beta> \frac{1}{2},
\alpha=\frac{1}{2}$ there is a continuous transition (i.e.\ one
accompanied by diverging lengthscales) to a phase in which the
particle current is independent of $\alpha$ and $\beta$.  The phase
diagram for the model is shown in figure~\ref{ASEP-phase}.

\begin{figure}
\begin{center}
\includegraphics[scale=0.5]{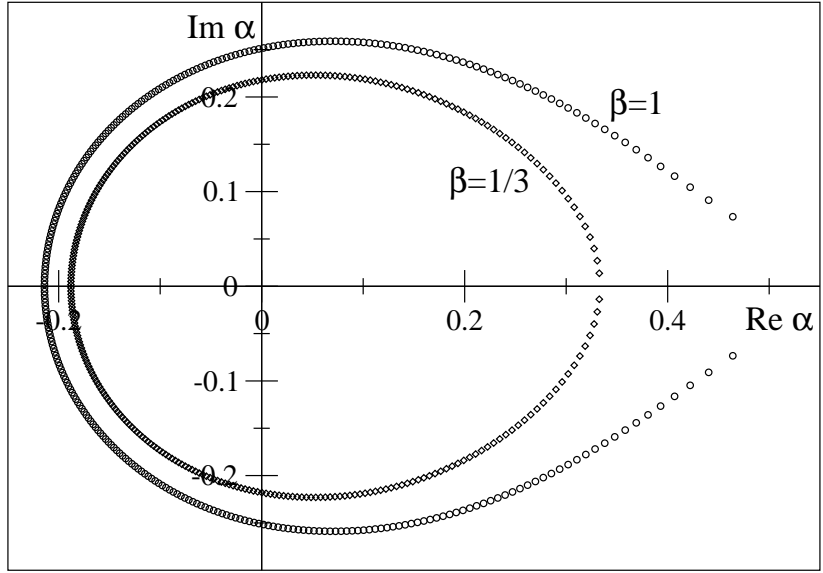}
\end{center}
\caption{\label{ASEP-alphazeros} Zeros of the normalisation in the
ASEP in the complex-$\alpha$ plane and $\beta=\frac{1}{3},1$ and
lattice size $N=300$.}
\end{figure}

The steady-state normalisation (\ref{Zneqm}) for the ASEP with $N$
sites has been calculated \cite{DEHP93} as
\begin{equation}
\label{Zasep}
Z_N(\alpha,\beta) = \sum_{p=1}^{N} \frac{p(2N-1-p)!}{N!(N-p)!}
\frac{(1/\beta)^{p+1}-(1/\alpha)^{p+1}}{(1/\beta) - (1/\alpha)} \;.
\end{equation}
It is a simple matter to use a computer algebra package to solve this
equation for its zeros in the complex-$\alpha$ plane at fixed $N$ and
$\beta$.  (Equivalently, one could look at the complex-$\beta$ zeros
at fixed $\alpha$ since $Z_N(\alpha,\beta) = Z_N(\beta,\alpha)$.)  In
figure~\ref{ASEP-alphazeros} plots of the zeros are shown for
$\beta=1$, where a continuous transition occurs at
$\alpha=\frac{1}{2}$, and for $\beta=\frac{1}{3}$, where a first order
transition occurs at $\alpha=\frac{1}{3}$.  We immediately notice that
the curve of zeros seem to intersect the real positive $\alpha$ axis
at the correct transition point.  Furthermore, the density of zeros
near the first-order transition point ($\alpha=\beta=\frac{1}{3}$)
seems to be uniform and nonzero, whereas the density of zeros near the
second-order transition point ($\alpha=\frac{1}{2}, \beta=1$) seems to
decrease to zero.  Both of these observations are in accord with the
results known for equilibrium partition function zeros discussed in
section~\ref{eqm}.

As shown in \cite{BE02}, the distribution of zeros in the
thermodynamic limit can be calculated once one knows that, for large
$N$, the normalisation behaves as $Z_N \sim A J^{-N} N^\gamma$.  In
this expression, $A, J$ and $\gamma$ depend on $\alpha$ and $\beta$
and the quantity $J$ is the current of particles across the lattice.
Thus, the complex free energy (\ref{hdef}) in the limit $N\to\infty$
is
\begin{equation}
\label{ASEPh}
h(\alpha,\beta) = \lim_{N\to\infty} \frac{\ln Z_N(\alpha,\beta)}{N} =
- \ln J(\alpha,\beta) \;.
\end{equation}
Furthermore for $\alpha$ smaller than the transition value $\alpha_c$,
$J=\alpha(1-\alpha)$ whereas for $\alpha>\alpha_c$,
$J=\alpha_c(1-\alpha_c)$.

Although an electostatic analogy was used in \cite{BE02} to find the
zero distribution, the mathematical content is the same as that used to
derive the two rules (\ref{boundary-rule}) and (\ref{density-rule}) in
section~\ref{eqm}.  Applying the first rule, which demands that the
real part of the free energy be continuous across a phase boundary, we
find that the zeros of $Z_N(\alpha,\beta)$ should lie along the curve
$|\alpha(1-\alpha)| =\alpha_c(1-\alpha_c)$.  It is therefore
convenient to change variable to $\xi=\alpha(1-\alpha)$.  Then, in the
complex-$\xi$ plane, the zeros lie on the circle $|\xi|=\xi_c =
\alpha_c(1-\alpha_c)$.  The density of zeros on this circle can be
found by setting $\xi = r \rme^{i\theta}$ and parametrising points on
the circle as $s = \xi_c \theta$.  Then, the second rule
(\ref{density-rule}) gives the density of zeros $\mu(s)$ on the circle
as
\begin{equation}
\mu(s) = \frac{1}{2\pi} \frac{\D}{\D s} \left[ \Im \ln \xi - \Im \ln
\xi_c \right] = \frac{1}{2\pi\xi_c} \frac{\D}{\D \theta} \theta =
\frac{1}{2\pi\xi_c} \;.
\end{equation}
That is, in the complex-$\xi$ plane the zeros should become evenly
distributed on a circle in the thermodynamic limit.  Transforming the
zeros of (\ref{Zasep}) obtained at different system sizes to the
complex-$\xi$ plane reveals this to be the case \cite{BE02}.

Finally, one can show that near the intersection point between the
curve of zeros in the complex-$\alpha$ plane and the positive real
$\alpha$ axis, the zeros of (\ref{Zasep}) sit on the curve $x =
\frac{1}{2} - (y^2 + \frac{1}{4} - \xi_c)^{1/2}$ where $x$ and $y$ are
the real and imaginary parts of $\alpha$.  For the case $\beta <
\frac{1}{2}$, $\xi_c < \frac{1}{4}$ and the transition is first order.
One finds that the curve of zeros is smooth at the transition point
$\alpha=\beta$, and that the density of zeros is $(1-2\beta)/[2\pi
\beta(1-\beta)]$ which is nonzero.  These are precisely the properties
of the equilibrium partition function zeros at a first-order
transition point (see section~\ref{eqm}).

At the continuous transition point ($\beta \ge \frac{1}{2}$ and
$\xi=\frac{1}{4}$, the zeros pass through the real $\alpha$ axis along
the line $x=\frac{1}{2} - |y|$.  Recall from section~\ref{eqm} that a
transition of $n^\th$ order has the curve of zeros meeting the
positive real axis at an angle of $\frac{\pi}{2n}$.  Here the zeros
clearly approach at an angle $\frac{\pi}{4}$ suggesting a second-order
transition.  In fact, one finds the density of zeros is $\frac{4}{\pi}
s$ at a distance $s$ along this line, confirming that the transition
is second-order.

In summary,  we have found that the Lee-Yang theory of
first-order and continuous phase transitions applies to the
normalisation of the nonequilibrium asymmetric exclusion process just
as it does to the partition function of equilibrium systems.  Of
course, this does not prove that the theory is generally applicable,
and so there is some value in investigating other nonequilibrium
steady states that exhibit phase transitions.

\begin{figure}
\begin{center}
\includegraphics[scale=0.75]{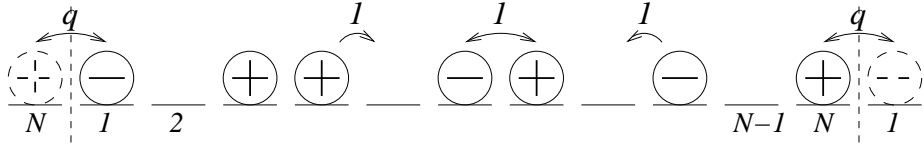}
\end{center}
\caption{\label{AHR} The dynamics defining the AHR model: labels
indicate the rates at which the hops denoted by arrows can occur.
Note that the $N$-site lattice is periodic.}
\end{figure}

One such state is that initially studied by Arndt, Heinzel and
Rittenberg \cite{AHR98,AHR99}.  This model comprises $M_{+}$ ($M_{-}$)
positively (negatively) `charged' particles on a closed ring of $N$
sites.  When next to vacant sites, the positive particles hop to the
right and negative particles to the left at unit rate, with hops in
the opposite directions disallowed.  Meanwhile, should two oppositely
charged particles be next to one another, the following transitions
can occur
\[ + - \stackrel{q}{\to} - + \qquad - + \stackrel{1}{\to} + - \]
where the label above the arrow indicates the rate at which the
transitions occur.  These dynamics are illustrated in figure~\ref{AHR}.

Although a matrix product solution for this model is known to exist
\cite{AHR99}, it is technically difficult to work out the solution in
its full generality \cite{RSS00}. However in the case of either a
single vacancy \cite{sasamoto00} or a single negative particle, i.e.\
$M_{-}=1$ \cite{Jafarpour00} the steady-state normalisation is known
explcitly and phase transitions have been identified.
In the case $M_{-}=1$ the zeros of the partition function in the
complex-$q$ plane have been studied \cite{Jafarpour02} and, once again
the continuous phase transition present in this model is correctly
described by the Lee-Yang theory.

A slightly different Lee-Yang analysis, that precedes the work
described above, has been used to study the nonequilbrium phase
transition that occurs at a finite density of vacancies
\cite{Arndt00}.  The motivation behind this study arose from computer
simulations of the model \cite{AHR98,AHR99} for the case where the
numbers of positive and negative particles were equal.  In these
simulations, three phases were initially identified: a `solid' phase
for small $q$, a `fluid' phase for large $q$ and a mixed phase
comprising the background `fluid' and a large mobile droplet in an
intermediate regime $1<q<q_{c}$ where $q_{c}$ is some
density-dependent quantity.

The steady-state normalisation $Z_{N,M}$ for the case of
$M_{+}=M_{-}=M$ positive and negative particles on the $N$-site ring
is not known exactly (at least in the canonical ensemble), therefore
Arndt \cite{Arndt00} chose to study the generating function
\begin{equation}
\label{AHRgf}
\Zed_N(z,q) = \sum_M z^M Z_{N,M}(q)
\end{equation}
in the complex-$z$ plane.  Physically this approach is equivalent to
placing the ring with nonequilibrium interactions in chemical
equilibrium with a particle reservoir at fixed fugacity $z$.  One can
check that as one takes the thermodynamic limit, the relative size of
the fluctuations in the density $r=M/N$ vanish, and so working at
fixed fugacity $z$ is equivalent to working at fixed density $r$. (In
fact, this is a standard `trick' for dealing with closed systems in
which the particle number is conserved, see e.g.\ \cite{DJLS93}.)

The key point here, however, is that $z$ is an equilibrium fugacity,
and not a microscopic transition rate, and so the Lee-Yang theory of
phase transitions described in section~\ref{eqm} ought to apply
directly here, without reference to the discussion of
section~\ref{NSS}.  It was found \cite{Arndt00} that when $q<q_{c}$
the zeros of (\ref{AHRgf}) in the complex-$z$ plane appear to lie
along ellipses, intersecting the positive real axis.  Conversely, for
larger $q$ the zeros appear to describe hyperbol\ae\ that avoid the
positive real axis.  After investigating numerically the density of
zeros in the complex-$z$ plane, Arndt concluded that there is a
fluid-fluid phase transition at some density-dependent point
$q_{c}>1$.

Unfortunately, an exact asymptotic (i.e.\ large $N$) analysis of the
generating function (\ref{AHRgf}) shows that the only nonanalyticity
occurs at the solid-fluid transition point $q=1$ \cite{RSS00}.
However it was noted that physical quantities vary rapidly, but
continuously, at some a point $q>1$. This phenomenom has been
explained as an abrupt increase in a corrleation length to an
anomalously large, but finite value \cite{KLMT02}.  We believe that
resolution with the study of Arndt lies in the fact that the system
sizes considered in \cite{Arndt00} were quite small ($N \le 100$),
whereas it was suggested \cite{RSS00} that the distinction between
this crossover behaviour and a genuine phase transition might become
apparant only once $N$ is increased above about $10^{70}$ (an
unfeasibly huge number!).

It would be interesting, therefore, to extend the numerical
computation of zeros performed in \cite{Arndt00} to much larger
systems, to see how the ellipses noted above develop.  For example it
might well happen that instead of approaching the positive-real fugacity
axis, the zeros of (\ref{AHRgf}) would terminate a short distance away
from it.

\subsection{Reaction-diffusion systems and directed percolation}
\label{RDandDP}

A large and important class of models with stochastic dynamics is
provided by reaction-diffusion systems \footnote{As an area of ongoing
research, new results are emerging continuously and there is no
up-to-date review that contains them all.  Nevertheless, one can
consult \cite{Redner97,Hinrichsen00} for an introduction.}.  In
contrast to driven diffusive systems, where the particle-particle
interactions imply conservation of particles, reaction-diffusion
systems are characterised by dynamics that result in a change in
particle number.  Moreover there are a number of such systems that
have absorbing states i.e.\ special configurations generally devoid of
particles that once entered cannot be left.  Phase transitions
associated with whether the system has a finite probability of not
being absorbed into such a state fall within the \textit{directed
percolation} and related universality classes \cite{Hinrichsen00}.  We
shall shortly discuss the second order phase transition associated
with the directed percolation university class in a little detail.
Meanwhile as a simple example of a reaction-diffusion system, we
review a model for which the steady state that can be solved using the
matrix product approach.  The approach again provides us with an
explicit expression for the normalisation (\ref{Zneqm}) by virtue of
which we can analyse its zeros in the plane of complex reaction rates.

The system in question \cite{HSP96} has for the dynamics at neighbouring
bulk sites on a one-dimensional lattice the processes
\[
\begin{array}{ccc@{\qquad}ccc}
\circ   \bullet & \stackrel{q}{\to}             & \bullet   \circ &
\bullet   \circ & \stackrel{q^{-1}}{\to}        & \circ   \bullet \\
\bullet \bullet & \stackrel{q}{\to}             & \bullet   \circ &
\bullet \bullet & \stackrel{q^{-1}}{\to}        & \circ   \bullet \\
\circ   \bullet & \stackrel{\kappa q}{\to}      & \bullet \bullet &
\bullet   \circ & \stackrel{\kappa q^{-1}}{\to} & \bullet \bullet
\end{array}
\]
occurring at the rates indicated, and with $\bullet$ representing a
particle and $\circ$ an empty lattice site.  It was demonstrated
\cite{HSP96} that the matrix product scheme used for the ASEP could be
generalised to cater for the steady state of the present
reaction-diffusion system on a lattice of $N$ sites with reflecting
boundary conditions (i.e.\ which is neither periodic nor has particle
input or removal at the boundaries).

In the matrix product approach, the normalisation $Z_N$ is given by a
scalar derived from $C^N$ where $C$ is a square matrix.  If $C$ is a
finite dimensional $m \times m$ matrix and can be diagonalised, the
resulting expression for $Z_N$ has the form
\begin{equation}
Z_N(q,\kappa) = \sum_{n=1}^{m} a_n \lambda_n^N
\end{equation}
in which $\lambda_n$ is the $n^\th$ eigenvalue of $C$ and the
coefficients $a_n$ arise from the details of the way in which one
obtains a scalar from the matrix $C^N$.  A normalisation of this form
leads to a complex free energy
\begin{equation}
\label{hmaxev}
h(q,\kappa) = \ln \left[ \max_n \{\lambda_n\} \right]
\end{equation}
in which the maximum means the eigenvalue with the largest
\textit{absolute} value.  Then, there is the possibility of a phase
boundary when the magnitude of the two largest eigenvalues of $C$ are
equal.  It turns out that for the closed reaction-diffusion system
introduced above, $C$ is a $4 \times 4$ matrix that can be
diagonalised when $q^2 \ne 1 + \kappa$ \cite{HSP96}. When $q^2 = 1 +
\kappa$, $C$ cannot be diagonalised on account of the largest
eigenvalue being degenerate, and a phase transition is found to occur
at this point.  Note that this scenario contrasts with the
transfer-matrix approach to one-dimensional equilibrium systems where
the partition function is also written as a product of matrices. Since
all elements of the transfer matrix are positive the largest
eigenvalue cannot become degenerate therefore there can be no phase
transition \cite{Evans00}.  However there is no such restriction on
the elements of $C$, and so eigenvalue crossing is permitted and
nonequilibrium one-dimensional phase transitions can occur.

The structure of the normalisation $Z_N(q,\kappa)$ is even simpler in
the case where particles are inserted onto and removed from the left
boundary at rates $\alpha$ and $\beta$ respectively (with the right
boundary remaining reflecting), and these rates satisfy the relation
$\alpha = \kappa(q^{-1} - q + \beta)$ \cite{Jafarpour03}.  In this
case $C$ is a $2\times 2$ matrix and, for $q^2 \ne 1+\kappa$, is
diagonalisable with eigenvalues
\begin{equation}
\lambda_1 = 1 + \kappa \quad\mbox{and}\quad \lambda_2 = q^2 \;.
\end{equation}
We see then immediately from (\ref{hmaxev}) and an application of the
rule (\ref{boundary-rule}) that there is a phase boundary in the
complex $q$ plane along the circle $|q| = \sqrt{1 + \kappa}$.  Also,
since one of the eigenvalues does not depend on $q$, the free energy
$h(q,\kappa)$ is a constant in one of the phases which, as the
analysis of the normalisation zeros for the ASEP demonstrated, implies
that the density of zeros on this circle is constant in the
thermodynamic limit.  In turn, this implies that the phase transition
is first order, as confirmed by explicitly calculating the density
profile in the two phases \cite{Jafarpour03}.

We finally turn our attention to a process that comprises symmetric
decoagulation (that is, $\bullet \circ \to \bullet \bullet$ and $\circ
\bullet \to \bullet \bullet$ taking place with equal probability in
each direction) and spontaneous decay ($\bullet \to \circ$) occurring
at independent rates.  Introduced as a crude model of an epidemic
\cite{Harris74}, this \textit{contact process} is known to exhibit a
transition from a phase in which the absorbing state (empty lattice)
is reached with certainty to a phase in which there is some
probability that the epidemic remains active forever in the
thermodynamic (infinite system size) limit \cite{Griffeath79}.  This
transition occurs as the ratio between the decoagulation and decay
rates is increased beyond a critical value.

\begin{figure}
\begin{center}
\includegraphics[scale=0.5]{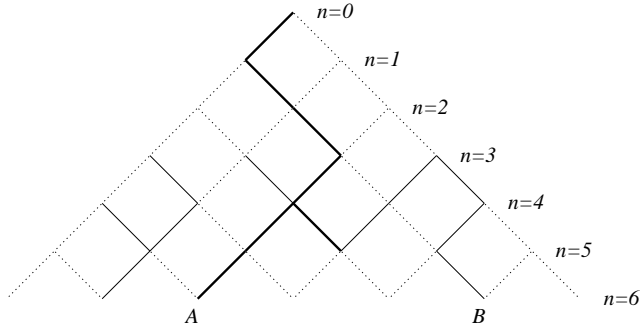}
\end{center}
\caption{\label{DP} Directed percolation.  Bonds are open (solid
lines) with probability $p$ and closed (broken lines) with probability
$1-p$.  A fluid starts at layer $n=0$ and may flow only downwards
through open bonds.  Bonds carrying the fluid are shown with thick
lines, so that the point $A$ on layer $n=6$ is connected to the origin
whereas point $B$ is not.}
\end{figure}

Although not proven, it is widely accepted on the basis of simulation
and approximate methods (such as series expansions) that the
transition just described is continuous and characterised by
\textit{directed percolation} (DP) exponents (see \cite{Hinrichsen00}
for an in-depth discussion of these issues).  Directed percolation
itself \cite{BH57} is a geometric construction designed to model fluid
flow through a random porous medium under the influence of gravity.
Consider the rhombic lattice shown in figure \ref{DP} and imagine
`pouring' fluid in from the uppermost site.  The fluid is allowed to
flow along a bond pointing diagonally downwards and then only if it is
open.  Each bond has a probability $p$ of being open (and hence a
probability $1-p$ of being blocked), and the state of each bond is
independent of any other and may not change with time.

The main quantity of interest in this system is the percolation
probability $P_n(p)$ that the fluid can penetrate to a depth $n$ in
the medium (see figure~\ref{DP}), averaged over all possible bond
configurations.  The order parameter for the model is the probability
$P_\infty(p)$ that the fluid can penetrate infinitely far given a bond
probability $p$.  If $p$ is below some critical percolation threshold
$p_c$, the fluid only ever penetrates a finite distance---i.e.\ a
layer becomes dry with certainty.  However, for larger $p$, the order
parameter becomes nonzero, growing as $P_\infty(p) \sim |p-p_c|^\beta$
with $\beta \approx 0.276$ near the critical point.  The importance of
the DP transition is that a wide range of models that have a
transition into an absorbing state are expected to have that
transition characterised by the DP exponents, one of which is $\beta$
\cite{Hinrichsen00}.  Despite a huge amount of interest in directed
percolation, none of the models expected to belong to its universality
class has been solved exactly.

Recently an attempt has been made to shed further light on the DP
transition by studying the zeros of the percolation probability
$P_n(p)$ \cite{ADH01,DDH02}.  At first glance, a connection between
this probability and a partition function is not obvious. However,
associating with each site a `spin' state $\sigma_i$, that is up if
site $i$ is connected to a point on the $n^\th$ layer and down
otherwise, yields a form for $P_n(p)$ that has the structure of a
transfer-matrix representation of a partition function for an
equilibrium system with three-spin interactions \cite{BG88}.

It is not clear that being able to give $P_n(p)$ the appearance of a
partition function necessarily implies that the Lee-Yang theory should
hold.  In particular, $P_n(p)$ has some features that set it apart
from `standard' equilibrium partition functions, in that it does not
grow exponentially in the number of bonds $N$, at least in the range
$0\le p\le1$.  It is also uncertain whether a free energy can
meaningfully be defined for this system, especially in the region $0
\le p \le p_c$ in which $\lim_{n\to \infty} P_n(p)=0$.

\begin{figure}
\begin{center}
\includegraphics[angle=-90,scale=0.5]{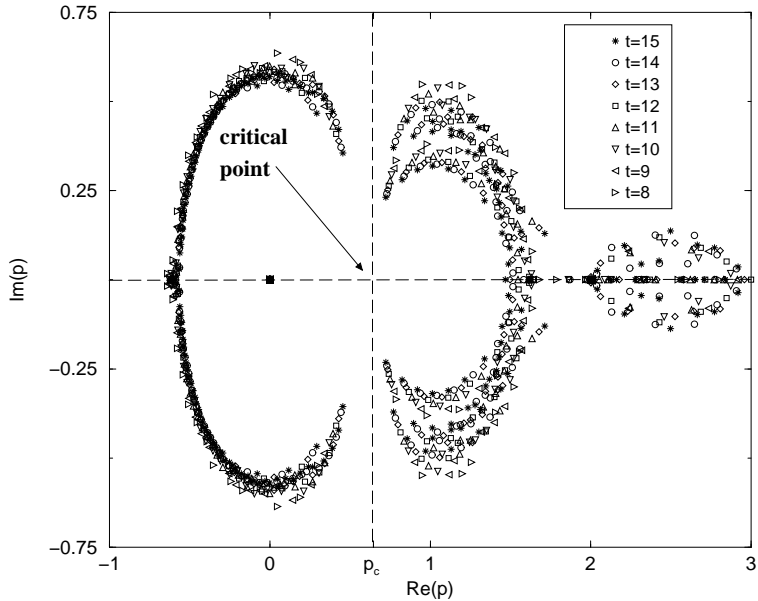}
\end{center}
\caption{\label{DDH-zeros} Zeros of the percolation probability on
the directed percolation lattice up to depths $n=15$.  Taken from
\cite{DDH02} in which the symbol $t$ is used where we use $n$.}
\end{figure}

Given these observations, it is perhaps not surprising that the zeros
of $P_n(p)$ calculated numerically for $n \le 15$ (leading to
polynomials of degree up to $N=240$) have a more complicated
distribution than for the (exactly solved) models considered so far.
As is evident from figure~\ref{DDH-zeros}, which was presented in
\cite{DDH02}, the zeros do not lie along a single curve but along a
sequence of curves meeting at the critical point (as well as inside
some region that encloses part of the real axis for $p>2$).

By considering the sequences of zeros approaching the positive real
axis, one can estimate the critical point $p_c$ and the density of
zeros as it is reached \cite{DDH02}.  The latter yields a prediction
for one of the DP critical exponents and one finds good agreement with
the most precise estimates of both the transition point and the
associated exponent obtained through other means.  Hence we have
further evidence for the applicability of Lee and Yang's ideas
concerning partition function zeros to a much wider range of
statistical distributions than equilibrium steady states.

\section{Summary and outlook}

In this work we have revisited the Lee-Yang description of equilbrium
phase transitions with a view to seeing whether the ideas apply to
more general nonequilibrium transitions.  Recently there have been a
number of studies of zeros of partition-function-like quantities that
arise in systems with nonequilibrium dynamics, and we have seen in our
review of these works that the Lee-Yang theory, as described in
section~\ref{eqm}, seems to hold quite generally.

We have argued that for dynamic models with a unique steady state, the
normalisation defined as a sum over the steady-state configurational
weights (\ref{Zneqm}) serves as a suitable `partition function' in the
sense that its zeros, in the complex plane of any model parameter,
should accumulate towards physical transition points in the
thermodynamic limit.  Furthermore, the density of zeros and angle of
approach to the real axis indicate whether the transition is
first-order (manifested physically through phase-coexistence) or
continuous (i.e.\ characterised by divergent correlation lengths and
times).  Thus studying the zeros of the normalisation (\ref{Zneqm})
provides an unambiguous classification of nonequilibrium phase
transitions as do the Lee-Yang zeros in the equilibrium case.

The observation that backs up this scenario is embodied by equation
(\ref{Zevprod}) which reveals that the reciprocal of the steady-state
normalisation is equal to the product of the characteristic relaxation
times in the dynamics.  Since near a phase transition one expects
timescales to diverge, one  also expects the normalisation to
approach zero.  However, a rigorous argument for this to be the case
is still lacking.  Moreover (\ref{Zevprod}) implies a possible link
between systems for which the steady state normalisation can be
calculated and those for which eigenvalues of the transition matrix
can in principle be calculated.

A different class of systems encompasses those whose steady state is
not unique.  The contact process is, in fact, an example of such a
model, in which the absorbing state is reached with certainty below
the critical decoagulation rate, whereas above it, and on an infinite
system, a second steady state can also be reached with some nonzero
probability.  Since this additional steady state exists only when the
lattice size becomes infinite, one must take that limit first, before
taking time to infinity.  Otherwise, on a finite system the steady
state is simply the absorbing state and the steady-state normalisation
is trivially equal to a constant.  Nevertheless the work of
\cite{ADH01,DDH02}, which we reviewed in section~\ref{RDandDP},
indicates that the Lee-Yang theory can be relevant when one considers
other properties of the nonequilibrium system such as the percolation
probability.

Although we have not discussed this in great detail here, it should be
noted that the Lee-Yang approach gives a method for extrapolating to
the thermodynamic limit from solutions for small system sizes.  In the
work of \cite{DDH02}, numerical solutions for small systems were used
successfully to estimate the transition point and density of zeros as
it is approached.  From this information one learns about the nature
of the phase transition and, for example, can estimate the values of
critical exponents.  This is a common technique in equilibrium
statistical physics (see e.g.\ \cite{JK02}) and it may be the case
that stochastic processes unyielding to analytical treatment could be
understood this way.

\begin{acknowledgments}

We thank Bernard Derrida for helpful discussions.
R.A.B.\ acknowledges financial support under EPSRC grant GR/R53197.

\end{acknowledgments}

\bibliographystyle{rabcite}

\end{document}